\documentclass[structabstract]{aa}
\usepackage{graphics}
\usepackage{natbib} 
\bibpunct{(}{)}{;}{a}{}{,} % to follow the A&A style 
\usepackage{float}
\usepackage{fixltx2e}

\begin{document}

\title{The detection of an older population in the Magellanic Bridge}

\author{G. Bagheri\inst{1}, M.-R.L. Cioni\inst{1, 2,}\thanks{Research
Fellow of the Alexander von Humboldt Foundation} and
R. Napiwotzki\inst{1}}

\offprints{g.bagheri@herts.ac.uk}

\institute{Centre for Astrophysics Research, College Lane, Hatfield
AL10 9AB, United Kingdom \and University Observatory Munich,
Scheinerstrasse 1, 81679 M\"{u}nchen, Germany}

\date{Received 2011 October 11/ Accepted 2013 January 17}

\titlerunning{Magellanic Bridge}

\authorrunning{Bagheri et al.}

\abstract {The Magellanic system comprises the Large Magellanic
Cloud (LMC), the Small Magellanic Cloud (SMC), and the less frequently
observed Magellanic Bridge and Magellanic Stream. The Bridge is traced
by neutral gas and has an observed stellar component, while the Stream
consists of gas only, with no observed stellar counterpart to
date.}{This study uses catalogues created in the direction of the Bridge from 2MASS and WISE to investigate the stellar content of the
Magellanic Bridge.}{Catalogues were created and colour-magnitude and two colour diagrams were analysed. A study was
also carried out on removing the Galactic foreground population in the
direction of the Magellanic Bridge, which was an important
consideration due to the low stellar density within the Bridge.}{This
study finds that the Magellanic Bridge contains a candidate older stellar population in addition to the younger population
already known.}{The formation of the Magellanic Bridge is likely to
have occurred from a tidal event between the LMC and SMC drawing most
of the material into it from the SMC. An older population in the
Bridge indicates that a stellar content was drawn in during its
formation together with a gas component.}


\maketitle

\section{Introduction}

The Magellanic Clouds are irregular dwarf galaxies within our Local
Group at a distance of $50-60$ kpc from the Milky Way \citep{mr00}. The Clouds consist
of the Large Magellanic Cloud (LMC) and the Small Magellanic Cloud
(SMC) which are connected by the Magellanic Bridge. The system also
contains the Magellanic Stream which is a gaseous feature between the
Clouds. It is widely believed that the Bridge was
formed tidally \citep{mat85}. The Magellanic Clouds are a prime template for studying galaxy
interactions, the derivation of their star formation history (SFH), and
the interaction between the Clouds and the Galaxy \citep{mr11}.

\begin{figure}
\resizebox{\hsize}{!}{\includegraphics{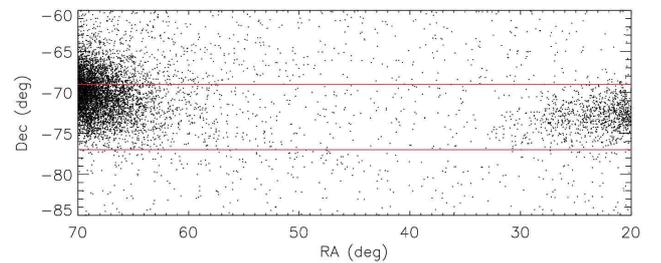}}
  \caption{Map of the stellar populations in the LMC, SMC and Bridge regions from 2MASS after the removal of Galactic foreground objects. The Bridge region of interest in this work is marked in red and spans from $20^\circ$ to $70^\circ$ in right ascension ($\alpha$) and from $-69^\circ$ to $-77^\circ$ in declination ($\delta$).}
\label{mag}
\end{figure}

The Magellanic Bridge spanning $\sim 13$ kpc (Fig. \ref{mag}) contains a known stellar
population \citep{ir90}. Star formation after a tidal event can be
studied by examining the ages, chemical abundances and kinematics. To date, only young populations have been
observed in the Bridge region, this implies that at the time of the
tidal interaction, only gas was drawn out of the Clouds to create the
Bridge, and that the inter-cloud population of the Bridge formed some
time afterwards. If an older population is present, it could
imply that stars as well as gas were stripped from the LMC and SMC in
the event that formed the Bridge. As tidal interactions should affect
both stars and gas, an older population is expected within the Bridge.

The aim of this study is to analyse near infra-red (NIR) public catalogues, to address the existence of an older inter-cloud population in the Magellanic
Bridge using 2MASS and WISE. This work will set a foundation for the analysis of catalogues from the VISTA Magellanic Clouds survey (VMC). An overview of the Magellanic Bridge is
presented in Sect. \ref{bridge}. Section \ref{data} describes the
2MASS and WISE data used in this study and how catalogues were created. A study of the Galactic foreground removal is carried out in Sect. \ref{fg}. The results of
the data analysis from each catalogue, including colour-magnitude diagrams (CMDs) and stellar populations
are shown in Sect. \ref{results}.  A discussion is given in Sect.
\ref{discussion} and the conclusions are summarised in
Sect. \ref{conclusions}.

\section{The Magellanic Bridge}
\label{bridge}

\citet{hin63}, carried out observations and digital reductions for a
survey of neutral hydrogen (\ion{H}{I}) in the direction of the Magellanic
Clouds and noted a considerable amount of gas between the two
Clouds. In a follow-up paper, \citet{hin263} concluded that the
Magellanic Clouds were encompassed by a halo of gas. They also
observed double peaks in the velocity of the SMC gas possibly
indicating two separate masses one behind the other.  \citet{denz73}
used observations of the 21cm \ion{H}{I} line to show a physical connection of
gas between the Clouds. This has been subsequently confirmed by other
authors and \citet{ir90} showed that the SMC Wing extends towards the
LMC into the Bridge. This is supported by the observations of
several regions within the Bridge containing young blue stars. Another
investigation by \citet{gron91}, using CCD photometry of three regions
within the Bridge recovered a population of young stars approximately
$10^{8}$ years old. \citet{gron92} observed two regions in the SMC
wing and one in the centre of the Bridge. They found that the SMC Wing
population has a large line of sight depth of $12\pm 8$ kpc and that
some associations are as young as $16$ Myr. This study also observed
that the inter-cloud stars were younger and less numerous than the
LMC, their distribution was not continuous, and were likely to have
formed from SMC gas.

\citet{mat85} proposed that the SMC had a close encounter with the LMC
warping the LMC disk and forming the Bridge while tidally fissioning
the SMC in a process of irreversible disintegration. It was also
postulated that the Clouds were not bound to the Galaxy and were
approaching from the direction of Andromeda. \citet{gron91} proposed a
scenario where a close encounter of the LMC and SMC occurred,
triggering star formation in the densest parts of the inter-cloud
region leaving dense gas on the SMC side.  Their analysis confirmed
that the Bridge associations on the west side formed first. It was
then suggested by \citet{gron92} that a tidal interaction between the
Clouds was on a milder scale than seen elsewhere in external galaxies,
but caused the formation of a disk around the SMC which was then
stripped to form the Bridge. A ring-like feature around the SMC has been suggested by \citet{mr06} and also \citet{ha04} corresponding to an older population of about $2.5$ Gyr.

Authors such as \cite{put99} and \cite{mull04} have supported the view that the Bridge may have been formed
from a close encounter $200$ Myr ago between the LMC and SMC, but this
is challenged with ideas that the Bridge was formed later due to the
difference between the Bridge and SMC populations and that the ages of
associations in the Bridge are closer to $6$ Myr. It is possible that
the Bridge was formed in a close encounter $200$ Myr ago but that star
formation was triggered by another event about $6$ Myr ago.

\citet{har07} used the MOSAIC II camera on the 4m telescope at
CTIO to observe twelve regions in the Magellanic Bridge searching for
an old stellar population which may have been stripped from the
SMC. The spatial distribution of these regions follows the \ion{H}{I} gas
ridge line in the Bridge and extends to right ascension $ \sim
3^\mathrm{h}$ where the \ion{H}{I} surface density falls to below the critical
threshold for star formation as determined by \citet{ken89}.  Harris
work showed no detection of an old population in the Bridge indicating
perhaps that the stars were all formed in-situ.

\section{Data}
\label{data}

The aim of this work is to establish source catalogues in the region
of the Magellanic Bridge prior to the analysis of observations using the VISTA
telescope \citep{em04} as part of the VMC survey \citep{mr11}. As such, the main source of data used in this work is
near-infrared (NIR) photometry from the Two Micron All Sky Survey
(2MASS). WISE data are also used for
comparison. The 2MASS and WISE catalogues are publicly
available.  The Bridge area was chosen from previous definitions of
the Bridge coverage by \citet{ker55}, \citet{mc81} and
\citet{har07}. This corresponds to an area between the LMC and SMC
from $70^\circ$ to $20^\circ$ in right ascension ($\alpha$) and from
$-69^\circ$ to $-77^\circ$ in declination ($\delta$). This is the first large scale study of the entire Bridge region, opposed to focussing on smaller observing regions.

2MASS collected raw imaging data covering $99.998$\% of the celestial
sphere in the NIR at $J$ ($1.25\mu$m), $H$ ($1.65\mu$m) and
$K_\mathrm{s}$ ($2.16\mu$m) between 1997 and 2001
\citep{sk06}. The point source detection level is $15.8$,
$15.1$, and $14.3$ mag at the $J$, $H$, and $K_\mathrm{s}$ bands
respectively, for the vast majority of observations. The $1\sigma$
photometric uncertainty is $<0.03$ mag for brighter sources above $K_\mathrm{s} = 12.0$ mag, with a pixel
size of $2.0^{\prime\prime}$. From this catalogue there were $\sim 300,000$ sources extracted in the Bridge area\footnote[1]{http://irsa.ipac.caltech.edu/Missions/2mass.html}.

The Wide-field Infrared Survey Explorer (WISE), \citep{wr10}, is an IR telescope that will survey the entire sky in the four WISE bands W1 ($3.4\mu$m), W2 ($4.6\mu$m), W3 ($12\mu$m) and W4 ($22\mu$m). The $5\sigma$ point source sensitivities are better than $0.08$, $0.11$, $1$ and $6$ mJy for the four bands respectively, with a point source detection level (Vega) of $16.5$, $15.5$, $11.2$ and $7.9$ mag. This corresponds to observations of a magnitude deeper than 2MASS $K_\mathrm{s}$ in $W1$ for objects with spectral types close to $A0$ stars and better for red sources. We extracted a total of $\sim 800,000$ sources in the Bridge field\footnote[2]{http://wise.ssl.berkeley.edu/index.html}.

The data for the Bridge region was downloaded from the 2MASS and WISE web interfaces using columns of $\alpha$ (deg), $\delta$ (deg),
magnitudes and photometric uncertainties with different flags used to
only extract stellar objects. In 2MASS flags were used to remove
objects with $JHK$ artefact contamination and/or confusion and
contamination from extended sources. WISE flags were applied to
remove multiple observations of sources, artefacts and maintain image and source
quality. Catalogues were then created by selecting only stellar
objects observed within each survey. The catalogues created are shown in Table \ref{tab} with the number of objects in each catalogue. The results in
this work are mostly affected by the photometric uncertainties in the
magnitudes. To remove objects with very large photometric errors close to the sensitivity limit, cuts at $K_\mathrm{s} = 14$ and $W2 = 14$ mag were applied. The photometric uncertainties of colours are derived from $\sigma_\mathrm{A-B} = \sqrt{\sigma_\mathrm{A}^{2} + \sigma_\mathrm{B}^{2} }$. The photometric errors on the magnitudes for each band are shown in Figs. \ref{errmag} and \ref{errmag2} for each catalogue along with cuts applied. Figure \ref{errmag} also shows that the photometric uncertainty increases for stars brighter than $J$, $H$ and $K_\mathrm{s} \sim 10.0$ mag due to saturation, and a cut at this level will also exclude Galactic giant stars as they will appear as very bright objects. A bright cut in WISE $W1$ and $W2$ bands at $8.0$ mag was also applied. Cuts at $14.0$ mag in $K_\mathrm{s}$ and $W2$ were chosen as a compromise between the higher photometric uncertainties near the sensitivity limit of the catalogue and not removing faint objects in the Magellanic Bridge. After the data had been cleaned, $\sim 87,000$ objects and $\sim 45,000$ objects remained in the 2MASS and WISE Bridge catalogues respectively.

\begin{table}[ht]
\caption{Number of stellar objects in each catalogue of the Magellanic
Bridge}   
\centering                          						
% used for centering table
\begin{tabular}{c c c}            						
% centered columns (3 columns)
\hline\hline                       						
Catalogue & Filter & Number of Bridge sources \\ [0.5ex]   		
\hline                             						
2MASS & $J$, $H$,$K_\mathrm{s}$ & $\sim 87,000$  \\ [1ex]                             						
WISE & $W1$, $W2$, $W3$, $W4$ & $\sim 45,000$ \\ [1ex]
\hline                              						
\end{tabular}
\label{tab}                          						
\end{table}

\section{Galactic foreground removal}
\label{fg}

Foreground Galactic stars towards low density regions of the
Magellanic System, like the Bridge, represent a non-negligible source
of contamination, as Galactic objects will appear with similar colours and magnitudes as Bridge objects, and will make up most of the
objects observed. The foreground population was
removed here using the different colours of objects within the Bridge and foreground regions, as well as magnitude cuts as described in Sect. \ref{data}. A target region was chosen to represent a purely Galactic
foreground population in the direction opposite to the Magellanic
Clouds across the Galactic plane, at the same Galactic longitude but at
positive Galactic latitude. This region was chosen as it represents a
similar location to the Clouds in terms of reddening effects and
foreground population, but without the galaxies dominating the
region it better represents a purely Galactic population. For this
work, catalogues were downloaded, centred on l$= 292^\circ$, b$= 39^\circ$, to cover an area
equivalent to that of the central Bridge region ($\sim 7^\circ \times
8^\circ$). 

Flags were used to clean up the catalogues as described in
Sect. \ref{data}. The foreground field was analysed for Galactic dust
reddening by examining dust extinction maps \citep{sc99}. The
average absorptions ($A_J=0.038$, $A_H=0.024$,
$A_K=0.015$) are sufficiently low to allow use of these fields in the
removal of Galactic foreground stars from the Magellanic Bridge
population and this is re-enforced as the Magellanic Clouds lie
sufficiently above the Galactic plane for reddening effects to be low
and therefore no correction for reddening has been made. It should be noted that the reddening
effects in the Bridge were also examined and the average absorptions
($A_J=0.043$, $A_H=0.028$, $A_K=0.018$) are also low
enough here and similar to the offset fields, to continue without a
reddening correction for the Bridge region.

\begin{figure*}
\resizebox{\hsize}{!}{\includegraphics{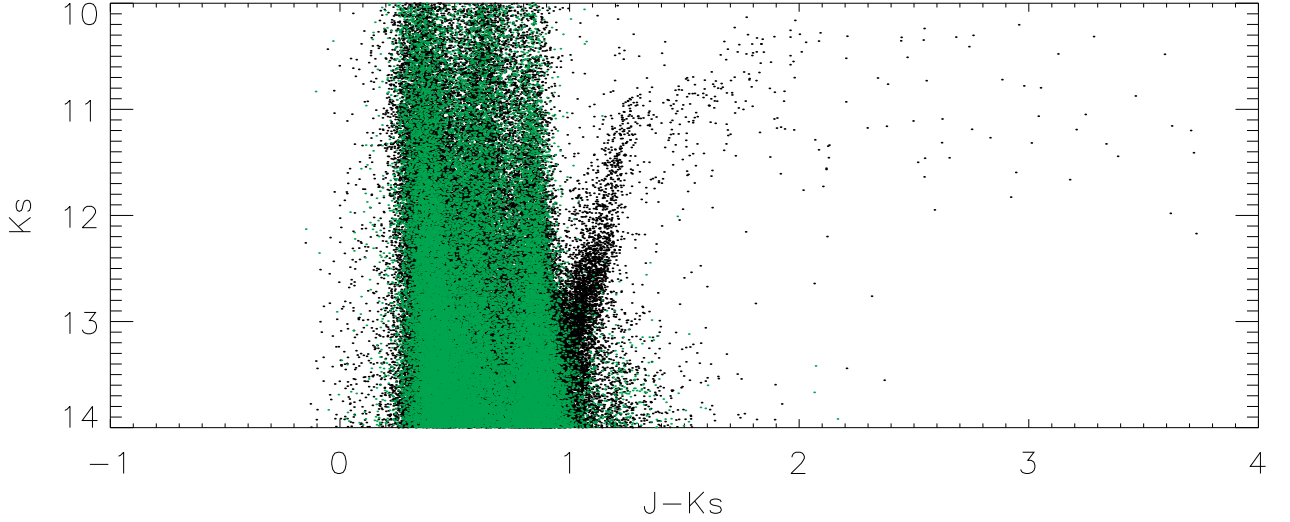}}
  \caption{CMD, $J-K_\mathrm{s}$ vs $K_\mathrm{s}$ of the Magellanic Bridge before the removal of Galactic foreground objects (black), and also the chosen Galactic foreground field (green) prior to the removal of foreground objects.}
\label{2masscmd}
\end{figure*}

Two approaches were applied to the removal of Galactic contamination from the Bridge sample. The first method is a more statistical approach, which does not supply information about the individual candidates remaining in the Bridge, but confirms that an over density of objects exists after the removal of the foreground. The second method makes use of a colour-cut foreground removal which, will remove some genuine Bridge objects, but will leave a sample of Bridge candidates that are almost free of contamination. The second method will enable the characterisation of any remaining Bridge population.

In each catalogue of the Bridge and foreground, the colour-magnitude diagram (CMD) chosen for the removal was that with the largest baseline and separation of objects. These were $J-K_\mathrm{s}$ vs $K_\mathrm{s}$ in 2MASS, and the WISE removal was based on the shorter wavelengths $W1-W2$ vs $W2$ as these bands are more likely to reveal a stellar population in the direction of the Bridge, and have magnitude limits comparable to 2MASS or slightly better. Objects that had 2MASS and WISE colours bluer than $-0.5$ mag and redder than $3.0$ mag were also discarded as these objects are likely to be background objects. These objects could also belong to a young or intermediate-aged Bridge population however, as this study addresses an older population this is not a concern. The statistical removal was carried out by binning the 2MASS and WISE CMDs in colour and magnitude bins, and comparing the number of objects in each bin from the foreground and Bridge field in terms of probabilities. 

\subsubsection{Probability foreground removal}

This foreground removal method was investigated by calculating the probability of an object belonging to the Bridge region, $N_\mathrm{Bridge}$, given the number of objects in the foreground field, $N_\mathrm{Foreground}$, (probability = $\frac{N_\mathrm{Bridge}-N_\mathrm{Foreground}}{N_\mathrm{Bridge}}$) per colour and magnitude bin. The cut off for an object belonging to the Bridge was chosen to be $0.75$ as this represents the upper quartile.

\subsubsection{Using colour-cuts for foreground removal}

The colour-cut approach compares the Bridge field and foreground field using a range of CMDs and two colour diagrams to find the best distinctions between the Bridge and the Galactic foreground stars. Instead of binning the Bridge and foreground, this method makes the use of cuts to separate the Bridge and foreground populations.

\subsection{2MASS}
\label{one}

\begin{figure*}
\resizebox{\hsize}{!}{\includegraphics{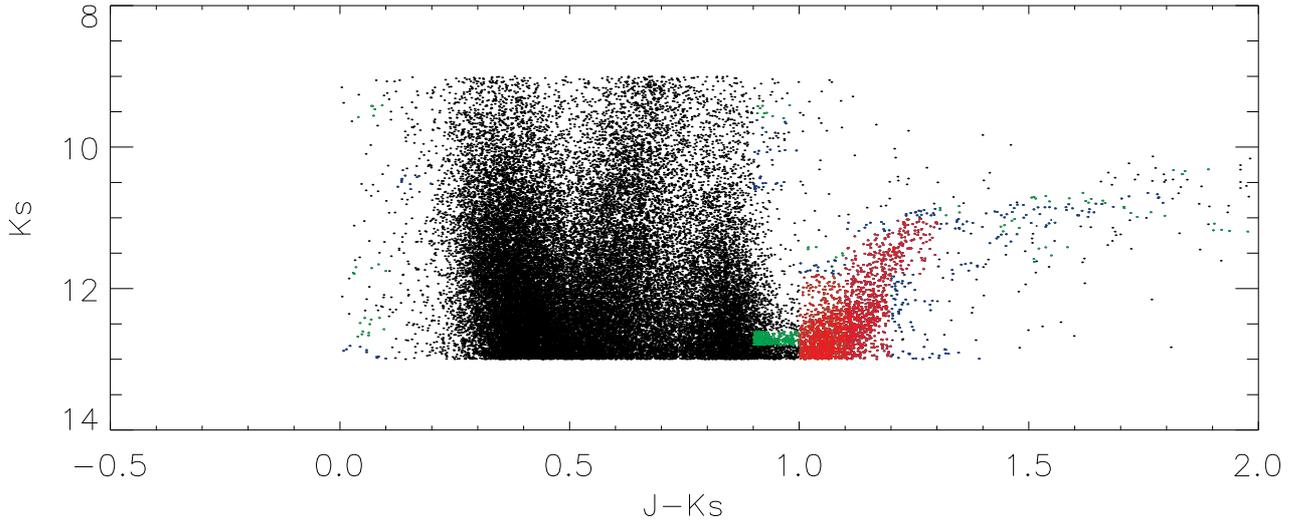}}
  \caption{CMD, $J-K_\mathrm{s}$ vs $K_\mathrm{s}$ of the Magellanic Bridge in 2MASS before the removal of Galactic foreground objects (black), and also after the application of the probability removal method, keeping only objects with probabilities of $0.75-0.85$ (green), $0.85-0.95$ (blue) and $0.95$ and over (red).}
\label{fgs}
\end{figure*}

The bin sizes chosen for the 2MASS statistical removal were $0.2 \times 0.5$ in colour and magnitude in all cases, as this bin size is large enough to contain a reasonable number of objects in each bin (up to $1000$ in this case), but small enough to have enough bins with different probabilities to optimise accuracy. From the 2MASS catalogue, taking the ratio of the total number of objects in the 2MASS Bridge catalogue vs foreground field, a value of P $> 0.75$ was used to select likely Bridge candidates with $2515$ objects remaining. This method confirms an over density of Bridge objects to the foreground and this method leaves a sequence remaining (Fig. \ref{fgs}) that is likely to belong to the Bridge given the position on the CMD (Sect. \ref{2mass}). 

The colour-cut method for the removal of foreground stars makes primarily use of the different IR colours of late type dwarfs and giants \citep{bes88}. For NIR 2MASS colours, the intrinsic colours of $H-K_\mathrm{s}$ and $J-H$ show a good separation between Galactic dwarfs and giants to Magellanic Bridge objects, enabling a good removal of foreground from the Bridge. The redder colours of the WISE $W1$ and $W2$ bands do not allow for such a clear separation since this region is dominated by the Rayleigh-Jeans tail towards redder wavelengths.

\begin{figure*}
\resizebox{\hsize}{!}{\includegraphics{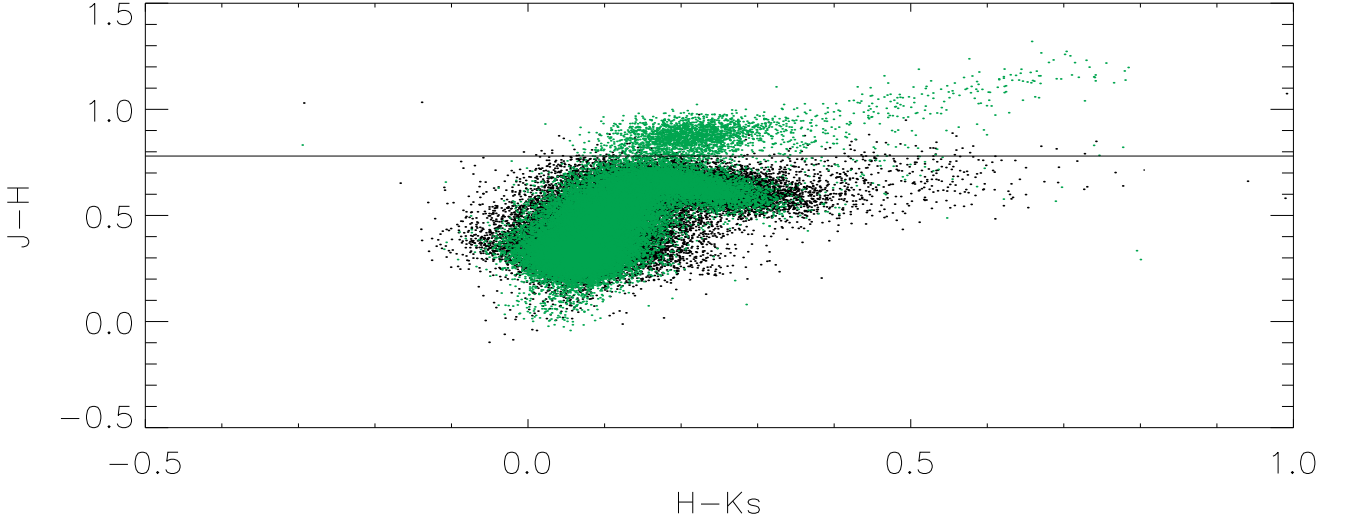}}
  \caption{Two colour diagram, $H-K_\mathrm{s}$ vs $J-H$ of the 2MASS Galactic foreground field (black) and the Magellanic Bridge (green), in order to apply a colour-cut method of removing the foreground objects. A line at $J-H > 0.78$ is shown to indicate the selection criteria that an object belongs to the Magellanic Bridge and not the Galactic foreground.}
\label{cuts}
\end{figure*}

Figure \ref{cuts} shows the two colour diagram $H-K_\mathrm{s}$ vs $J-H$ with the application of a cut at $J-H > 0.78$ to remove the Galactic objects from the 2MASS Bridge catalogue where there is a clear separation between Magellanic and Galactic objects. This colour selection criteria for the 2MASS-catalogue was applied to the 2MASS Bridge catalogue and from this colour-cut removal, $2499$ Bridge candidates remain. The Poisson error, $\sqrt{N} = 382$ for the Bridge catalogue, confirming that the candidates here are above statistical uncertainty.

\begin{figure*}
\resizebox{\hsize}{!}{\includegraphics{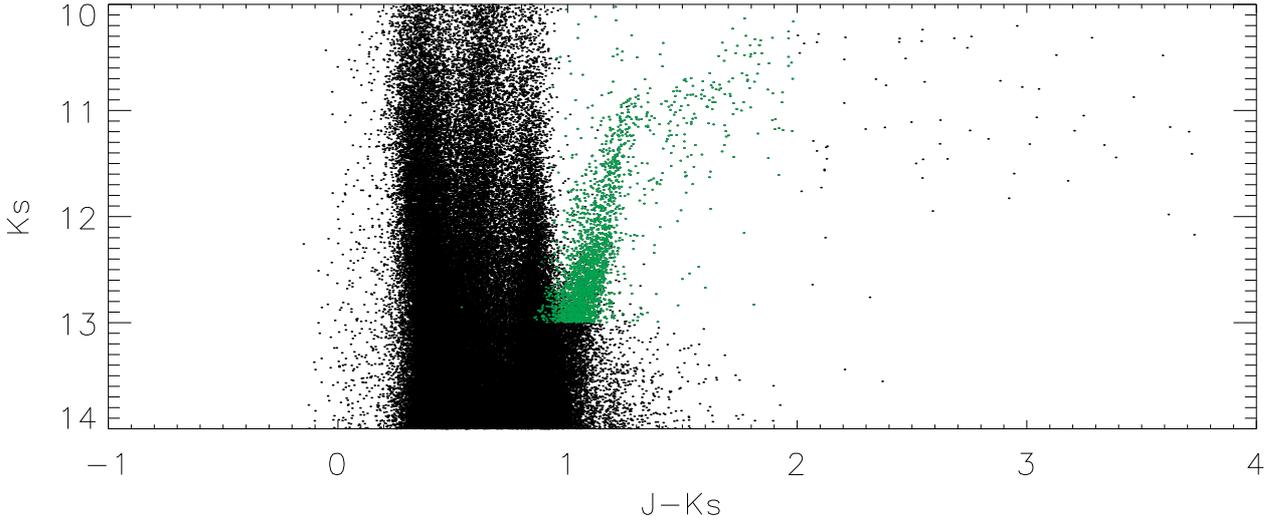}}
  \caption{CMD, $J-K_\mathrm{s}$ vs $K_\mathrm{s}$ of the Magellanic Bridge in 2MASS before the removal of Galactic foreground objects (black), and also after the application of the color-cut removal method (green).}
\label{fgs2}
\end{figure*}

After the application of the removal of Galactic foreground objects from the Bridge 2MASS catalogue, the statistical method confirm that there is an over density of Bridge objects to foreground contamination. The colour-cut method of removal isolates particular populations of Bridge candidates and these remaining objects will be investigated in Sect. \ref{results}.

\subsection{WISE}

The same methods of Galactic foreground removal were applied to the WISE catalogues as in 2MASS. The binning in WISE was applied at $W1-W2 = 0.1$ and $W2 = 0.2$ in order to maximise the number of objects in each bin (as in Sect. \ref{one}), while keeping the greatest accuracy. After the application of the probability method of removal to the WISE catalogue, just $75$ Bridge candidates remain, which is below the statistical threshold to confirm an over density within the Bridge. The colour-cut method of removing the Galactic foreground was investigated using the $W1$ and $W2$ WISE bands. 

\begin{figure*}
\resizebox{\hsize}{!}{\includegraphics{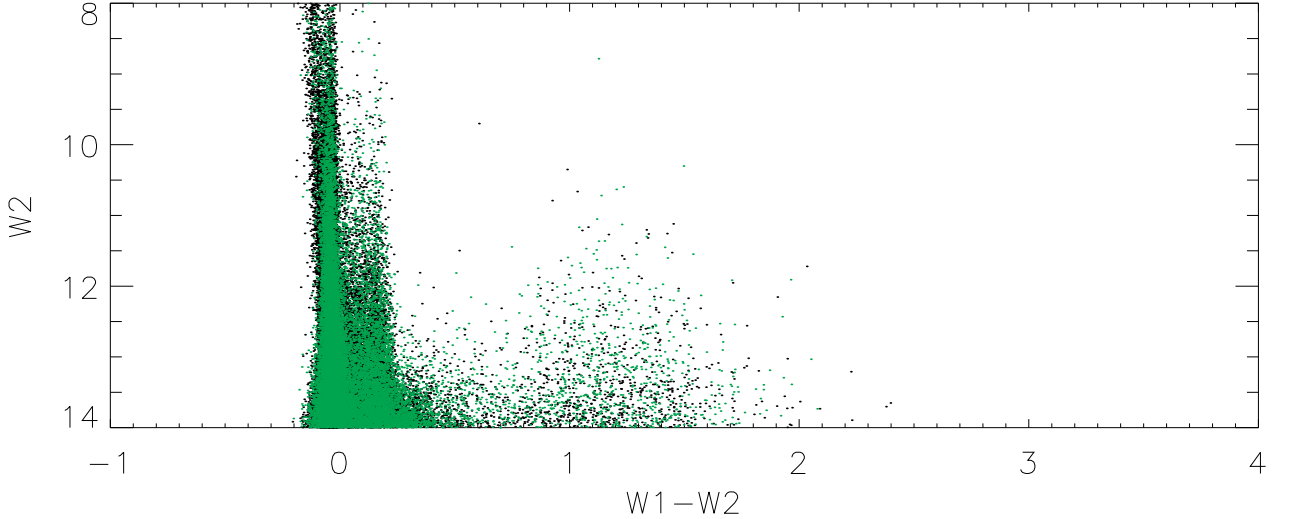}}
  \caption{CMD, $W1-W2$ vs $W2$ of the Magellanic Bridge in WISE (black), and the Galactic foreground field (green).}
\label{wisecmd}
\end{figure*}

It can be seen from Fig. \ref{wisecmd} that there is very little separation between the Bridge and foreground field in $W1-W2$ vs $W2$. This is also the case in other CMDs and two-colour diagrams meaning that it is not possible to separate the Bridge from Galactic objects using a colour-cut method. From the investigation of WISE at this point, there are no clear Bridge candidates to put forward for analysis. Due to the longer wavelengths of WISE, the catalogue is strongly dominated by background objects and the Rayleigh-Jeans regime of stellar objects, making it difficult to separate Bridge and foreground objects. The WISE objects may also be too red to produce a Bridge candidate population. The candidates from the 2MASS colour cut removal will be put forward for analysis, in order to characterise their populations.

\section{Stellar populations within the Bridge}
\label{results}

Thi work attempts to identify and characterise an older population within the Magellanic Bridge. The results are shown in the form of CMDs showing the different stellar populations making use of work by \citet{nik00}.

\subsection{2MASS}
\label{2mass}

In order to characterise the stellar populations within the Bridge from 2MASS,
a CMD was created in $(J-K_\mathrm{s})$ vs $K_\mathrm{s}$ with the application of boxes from
\citet{nik00}. These boxes were created to characterise the population
of the LMC, and have been adapted here for the Magellanic Bridge using the distance modulus at an average Bridge distance of $55$ kpc between the LMC \citep{wa12} and SMC \citep{gra12}. The boxes were then overlaid onto the observed CMD of the Bridge.

\begin{figure*}
\resizebox{\hsize}{!}{\includegraphics{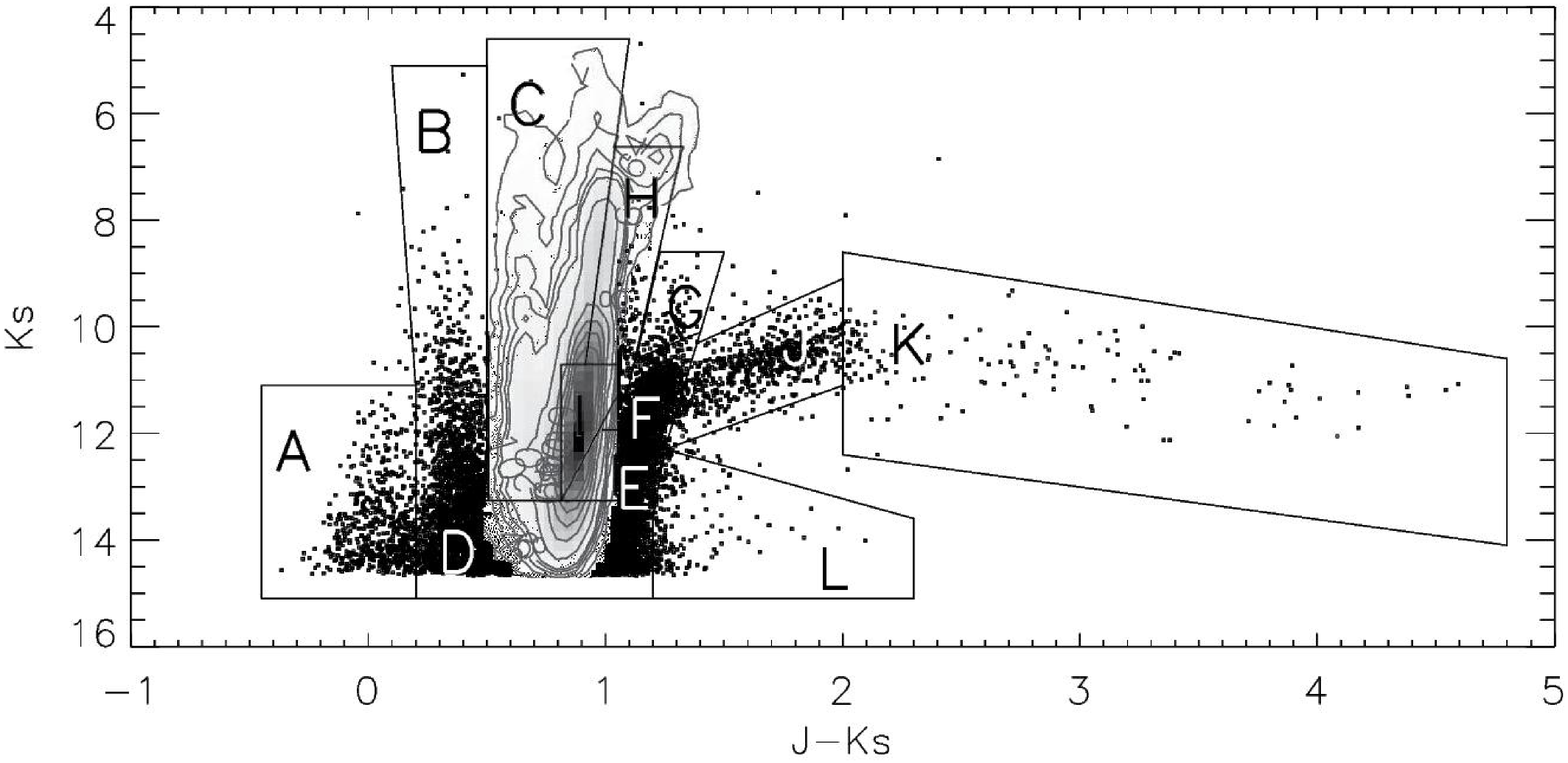}}
\resizebox{\hsize}{!}{\includegraphics{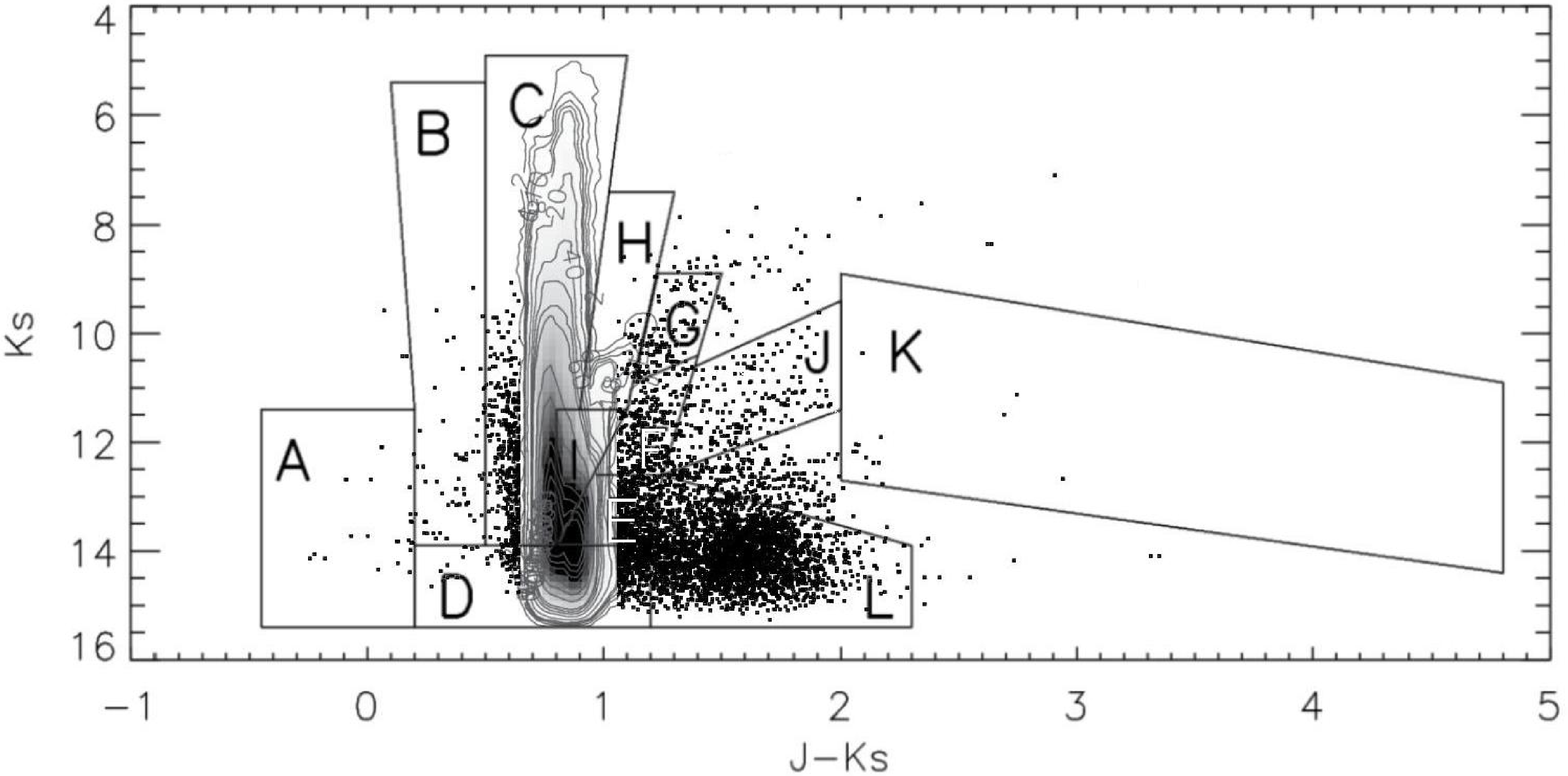}}
  \caption{Hess diagram, $(J-K_\mathrm{s})$ vs $K_\mathrm{s}$, of the LMC (top), and 
Magellanic Bridge (bottom) from the 2MASS catalogue. Boxes overlaid on the plot
are taken from \citet{nik00} and foreground stars have not been removed from the
catalogues. Here for the Bridge, the boxes have been shifted by 0.25 mag in
$K_\mathrm{s}$. Contour levels are at: $10$, $20$, $30$, $40$, $50$ and $60$ stars.}
\label{2MDBRG}
\end{figure*}

\begin{figure*}
\resizebox{\hsize}{!}{\includegraphics{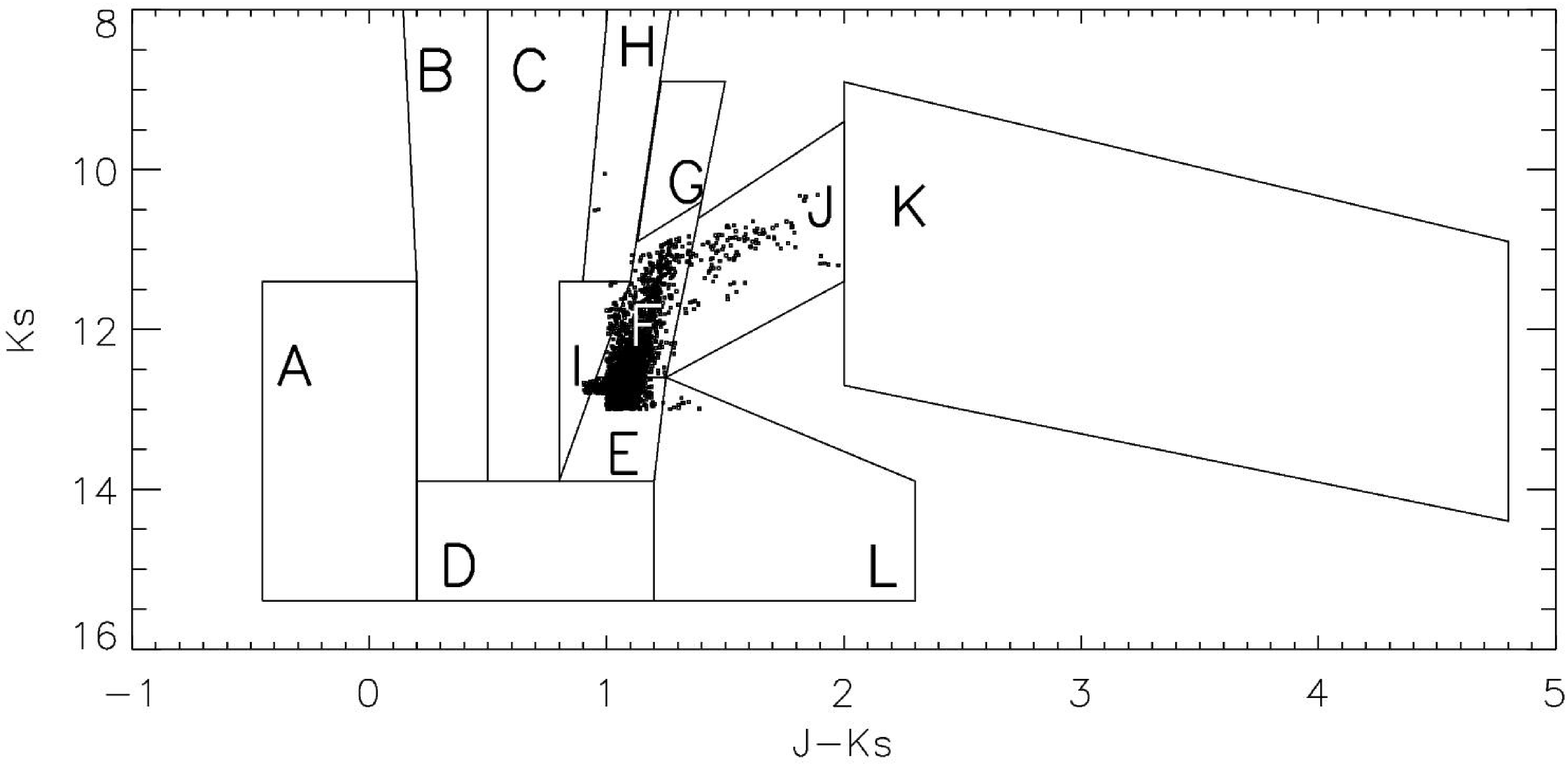}}
  \caption{CMD, $(J-K_\mathrm{s})$ vs $K_\mathrm{s}$ of Bridge stars from 2MASS observations after the removal of Galactic foreground stars. The boxes overlaid on the plot
are taken from \citet{nik00}. The boxes have been shifted by $0.25$ mag for the Bridge to account for the distance difference.}
\label{2MDBRGfg}
\end{figure*}

There are twelve regions identified  marked from A
through L and several techniques were applied by Nikolaev \& Weinberg
to identify the populations in each region. The LMC populations were
identified based on the NIR photometry of known populations from the
literature. The authors also performed a preliminary isochrone
analysis, where they matched the features of the CMD with
\citet{leo00} isochrones, to derive the ages of populations and draw
evolutionary connections among the CMD regions. The regions of the CMD
present in an outer region of the LMC, and the Bridge population from the 2MASS catalogue prior to the Galactic foreground removal are presented in Fig. \ref{2MDBRG}. The characterisation of the central Bridge population after the removal of Galactic foreground (Fig. \ref{2MDBRGfg}), are primarily E, F, and J, with a small contribution from L and I. Even considering photometric uncertainties in magnitudes and colour, the stellar population still fits well into the boxes characterising the different types of stars.\\

A summary of the types of stars found in the regions according to
Nikolaev \& Weinberg:

Region E covers the upper red giant branch (RGB) and includes the tip of the RGB. The tip
of the RGB is defined by the ignition of the degenerate helium burning
core in old (low mass) stars, the helium flash \citep{ren88}. Stars,
after ignition, evolve rapidly to the horizontal branch.

Region F contains primarily oxygen-rich asymptotic giant branch (AGB) stars of intermediate age
that are the descendants of stars in region E. Some fraction of region
F stars are long-period variables and reddened supergiants.

Region J sources are primarily carbon rich thermally pulsing (TP)-AGB stars. These stars
are descendants of oxygen-rich TP-AGBs in regions F and G. Region K
contains extremely red objects identified as obscured AGB carbon-rich
stars, as well as some dusty oxygen-rich AGB stars. Their large $(J-K_\mathrm{s})$ colours are due to dust in their
circumstellar envelopes.

The population in region I is young, with an age of approximately
$500$ Myr. A significant contributor to the source density in this
region are K- and M-type supergiants.

From the above population analysis, it can be seen that the youngest populations remaining within the Bridge after the removal of Galactic foreground are a very small number of K and M supergiants with an age of $\sim 500$ Myr. There is also a negligible contamination of brighter objects in box H which represents K- and M-type giants. These objects are still expected to be older than the assumed age of the Bridge implying that the red objects here are all older than the postulated Bridge age of $\sim 200$ Myr. The majority of Bridge candidates remaining fall into the boxes representing an old/intermediate population of RGB and AGB stars with ages of $\sim 400$ Myr and older.

\subsection{SIMBAD objects within the Bridge}
\label{dist}

\begin{figure*} 
\resizebox{\hsize}{!}{\includegraphics{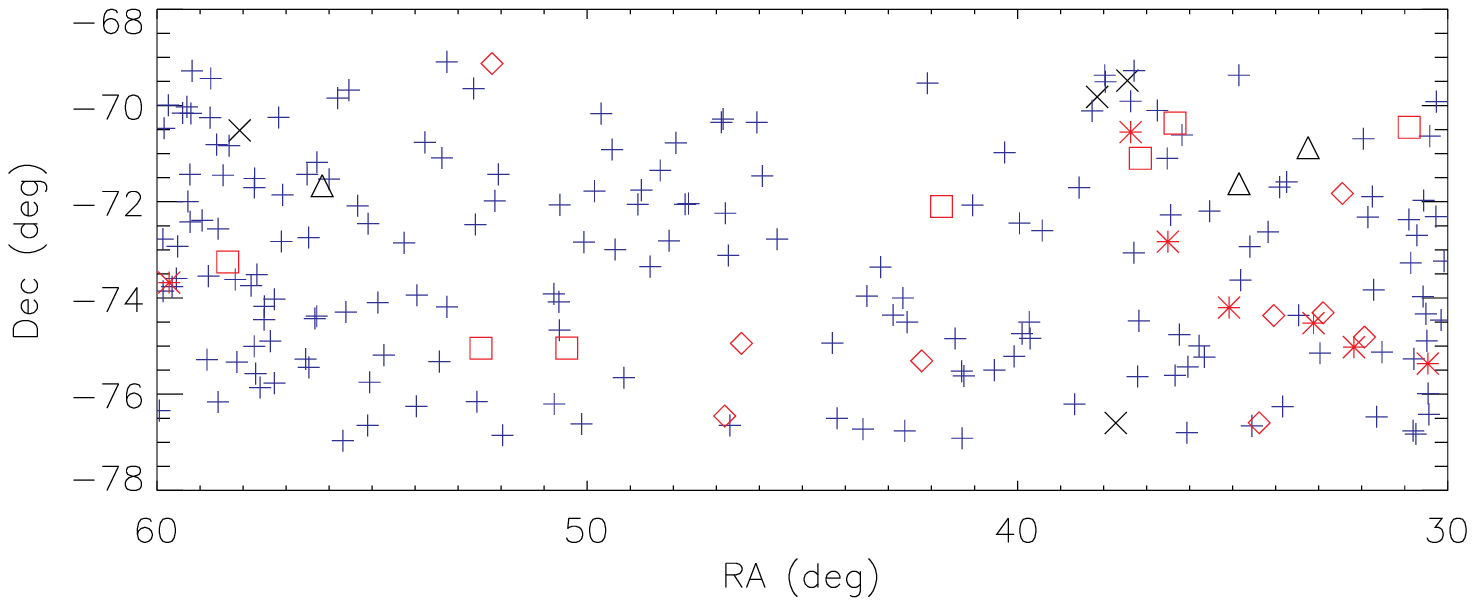}}
\resizebox{\hsize}{!}{\includegraphics{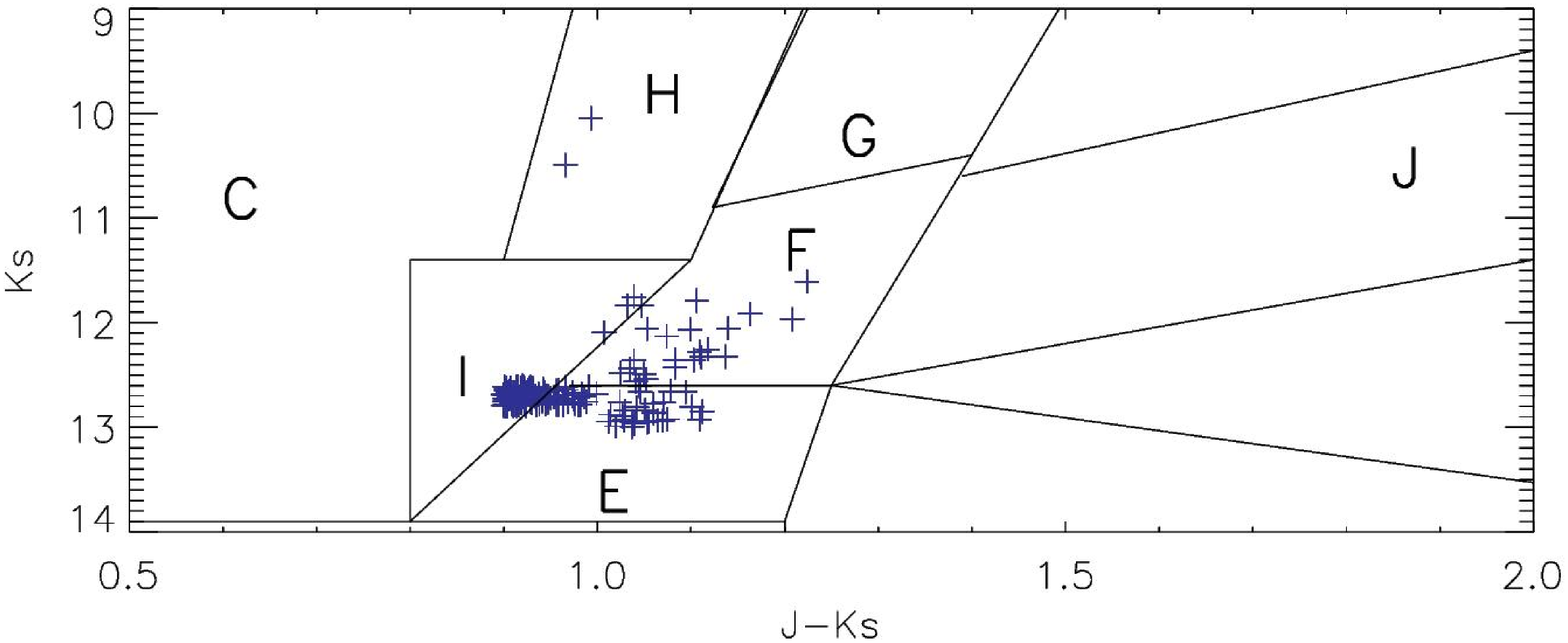}}
  \caption{The spatial distribution (top) of the different populations
present within the central Magellanic Bridge ($30^\circ$ to $60^\circ$) from the 2MASS catalogue after the
removal of Galactic foreground. The different
symbols represent: background galaxies (crosses), known SMC sources (red stars),  known
Bridge objects (red squares), known Magellanic
Cloud halo sources (red diamonds), other known sources (triangles) and new previously unpublished
sources (blue plusses) from SIMBAD. The $(J-K_\mathrm{s})$ vs ($K_\mathrm{s}$) CMD of unknown objects is also presented (bottom), with boxes overlaid as in Fig. \ref{2MDBRG}.}
\label{simbad}
\end{figure*}

The remaining objects in the 2MASS catalogue after the colour-cut removal of Galactic foreground, were
cross-matched to the SIMBAD database and were characterised in
$\alpha$ and $\delta$ (Fig. \ref{simbad}). Objects here belong to the centremost $30^\circ$ to $60^\circ$ of the Bridge to remove contamination from the LMC and SMC to isolate a true inter-Cloud population.

It can be seen from the lower panel of Fig. \ref{simbad} that the unknown, central Bridge candidate population is contained within boxes E, F and I representing an older Bridge population. Table \ref{2M} shows a breakdown of the objects as identified by
SIMBAD within the central Magellanic Bridge after the removal of Galactic
foreground and background contamination. It should be noted that the objects occupying region J from Fig. \ref{2MDBRGfg} are removed when focusing on the central Bridge, indicating that these dustier AGB stars are confined to the outer LMC and SMC regions. From the SIMBAD search, objects belonging to the SMC have been recovered, along with a small number of previously observed younger Bridge objects. A small number of objects remaining were background galaxies which have been removed from the final sample. The majority of objects have an unknown origin, which after the colour-cut removal of contamination, a number of these unknown objects are strong candidates for an older Bridge population.

\begin{table*}[ht]
\caption{Different populations identified within the Magellanic Bridge from
SIMBAD. [1] \citet{ir90}, [2] \citet{can18}, [3] \citet{hog98}, [4] \citet{sk06}, [5] \citet{hol74}, [6] \citet{ab95}, [7] \citet{west71}, [8] \citet{dem91}.}   
\centering                          						
% used for centering table
\begin{tabular}{c c c c}            						
% centered columns (4 columns)
\hline\hline                       						
Object & No. within 2MASS Bridge catalogue & Reference \\ [0.5ex]   		
\hline                             						
Bridge star & 7 & [1]\\ [1ex]                           						
SMC star & 7 & [1] [7]\\ [1ex]                              						                            						
Magellanic Cloud object & 9 & [1] [8]\\[1ex]                             						                          	
Background Galaxy & 4 & [4] [5]\\[1ex]                         						
Other known object & 5 & [2] [3] [6]\\ [1ex]                            						
Unknown & 202 &  \\ [1ex]
\hline                              						
\end{tabular}
\label{2M}                          						
\end{table*}

\section{Discussion}
\label{discussion}

\citet{har07} created an infrared CMD of a $35$ deg$^2$ region of the
central part of the Magellanic Bridge compared to a region of the
outer Bridge towards the LMC. He found that in the outer region of the
Bridge there was an abundance of RGB and AGB stars associated with the
LMC but in the central region there were hardly any of these
populations. The work here uses a method of Galactic foreground
removal using colour-cuts, and finds a population within the Bridge including RGB and AGB
stars within the Galactic foreground subtracted
stellar Bridge population.  The ages of the 2MASS RGB and
AGB populations identified here could range from $\sim 400$ Myr to $5$ Gyr for redder,
lower-luminosity stars \citep{fr90}. This indicates the presence of a
population older than the age of the Bridge that is likely to have
been drawn in tidally at the time of the Bridge forming event. An
alternative explanation could be that the Bridge is older than
previously estimated. \citet{ol11} have identified a population of SMC objects within the LMC. These objects have lower metallicity than the LMC but are consistent with SMC giants which supports the theory of tidal interaction. With the discoveries of LMC and SMC objects reaching these great distances, and the work presented here, it is possible that there is not a `bridge` between the two Clouds, but that each Cloud has a large halo that overlaps with the other. Metallicity measurements of Bridge objects will constrain their membership to either the LMC or SMC and will determine whether the Bridge objects belonging to each Cloud overlap or are separated.\\

Other recent studies \citep{ch10} have used mid-IR Spitzer IRAC and
MIPS observations to detect a number of young stellar objects (YSOs)
in the Magellanic Bridge, mostly nearby high \ion{H}{I} density regions. They
have also studied how star formation has proceeded and estimated the
star formation efficiency in the Bridge by looking for Herbig Ae/Be
stars using known catalogues of blue stars and stellar clusters. If
the new populations in this work are genuine Bridge members they will
contribute to the present knowledge of star formation within this
region.\\

The Magellanic Bridge is one tidal Bridge between two irregular
galaxies.  \citet{be08} investigated a Bridge that formed tidally
between the outer halos of M31 and M33 made only from \ion{H}{I} gas with no
stars associated to it. Its origin is also under debate and it could
be in-fact a tidal stream like the Magellanic Stream containing no or
a very small stellar component. \citet{sm10} have used the GALEX
ultraviolet telescope to study various tidal features including
bridges, spiral arms and tails of gas. There are many tidal features
in existence, some with and some without stars within the gas in galaxies out to distances of $50$ kpc \citep{pak04}. A more
detailed study is required to determine a model to distinguish a tidal
bridge from a stream, spiral arm or tail and a study into the
conditions that form these features will give more information to the
likely presence of stars within them. The Magellanic Bridge has had a
known stellar population however, in the Magellanic Stream no stars
have been detected within the \ion{H}{I} gas. Perhaps unless drawn in
at formation as indicated here, tidal features do not contain a stellar population until
an event occurs that is strong enough to trigger star formation within
the tidal feature in an environment with a high enough gas
density. Further observations of tidal systems, including the
Magellanic Clouds with VISTA, are needed to give us a better picture
of the interactions that form these features and star formation
within.\\

\section{Conclusions}
\label{conclusions}

To date, the majority of current
observations of the Magellanic Bridge have been restricted to the \ion{H}{I} ridge line and the
southern area. This is the first infrared study of the entire Bridge, which reveals candidates for an older population that were tidally stripped from the Magellanic Clouds as stars.
The results of the CMD analysis suggest that the Bridge may contain:
\vspace{-0.2cm}
\begin{enumerate}
  \item upper RGB stars to the tip of the RGB,
  \item RGB and E-AGB stars,
  \item Carbon stars, O-rich and dusty AGB stars,
  \item reddened M-type giants.
\end{enumerate}

These stars indicate ages older than previously suggested for the
Bridge. The ages of the RGB and AGB stars in the central Bridge
region is likely to range from $\sim 400$ Myr to $5$ Gyr meaning that these stars
were drawn into the Bridge at the formation event and did not form
in-situ. In order to
better constrain the above, follow up observations of the Bridge
populations are required as part of the VMC survey and also of
metallicities and radial velocities from spectra that may constrain
the membership of Bridge stars to the LMC or SMC.

\acknowledgements

{This publication makes use of data products from the Two Micron All
Sky Survey, which is a joint project of the University of
Massachusetts and the Infrared Processing and Analysis
Center/California Institute of Technology, funded by the National
Aeronautics and Space Administration and the National Science
Foundation, the Wide-field Infrared Survey Explorer, which is a joint project of the University of California, Los Angeles, and the Jet Propulsion Laboratory/California Institute of Technology, funded by the National Aeronautics and Space Administration and also the SIMBAD
database, operated at CDS, Strasbourg, France.}

\bibliographystyle{aa} 
\normalsize 
%\phantomsection 
\label{listofbibs}
%\addcontentsline{toc}{chapter}{Bibliography} 
\bibliography{bridge}

\appendix{}

\section{Figures}
\begin{figure}[H]
\resizebox{\hsize}{!}{\includegraphics{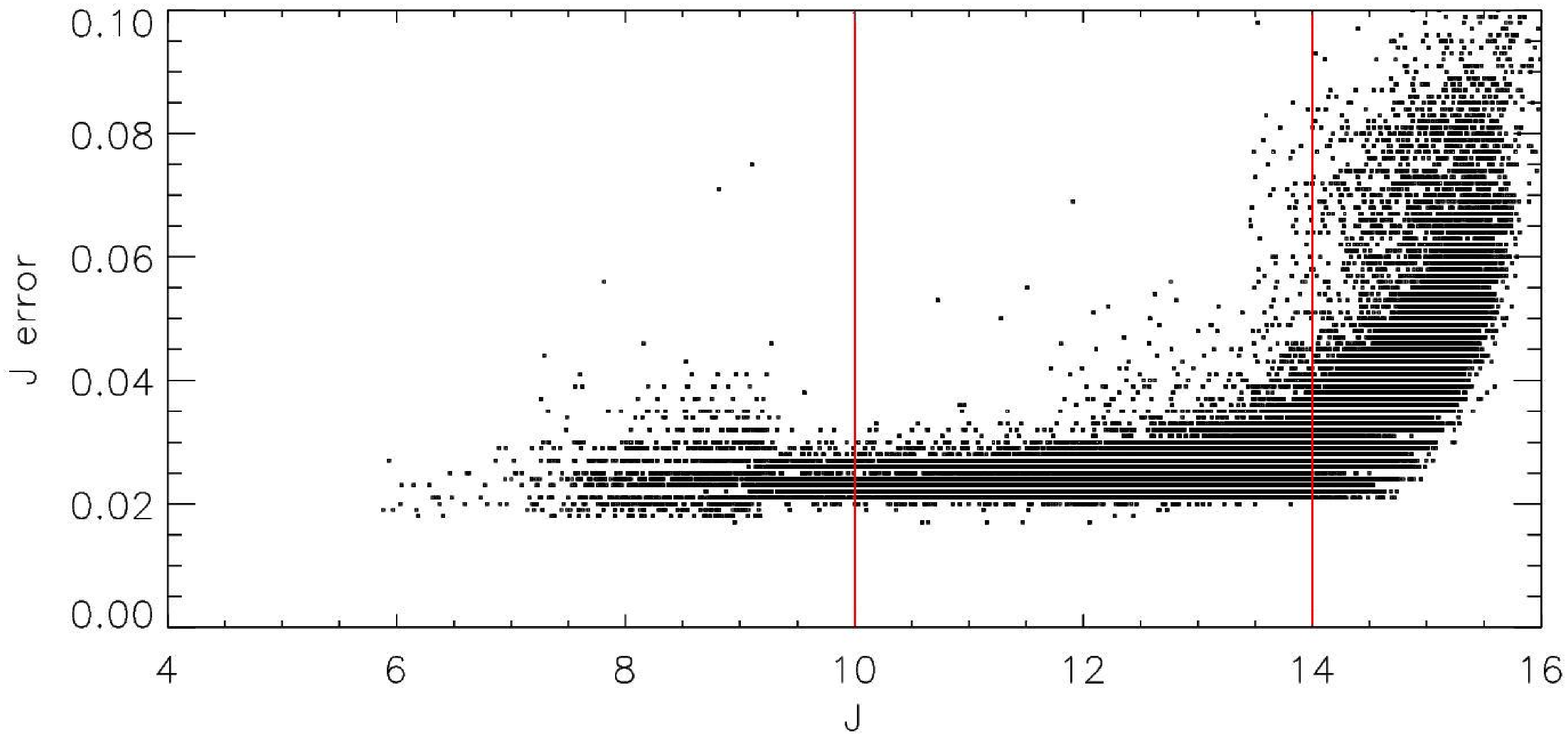}}
\resizebox{\hsize}{!}{\includegraphics{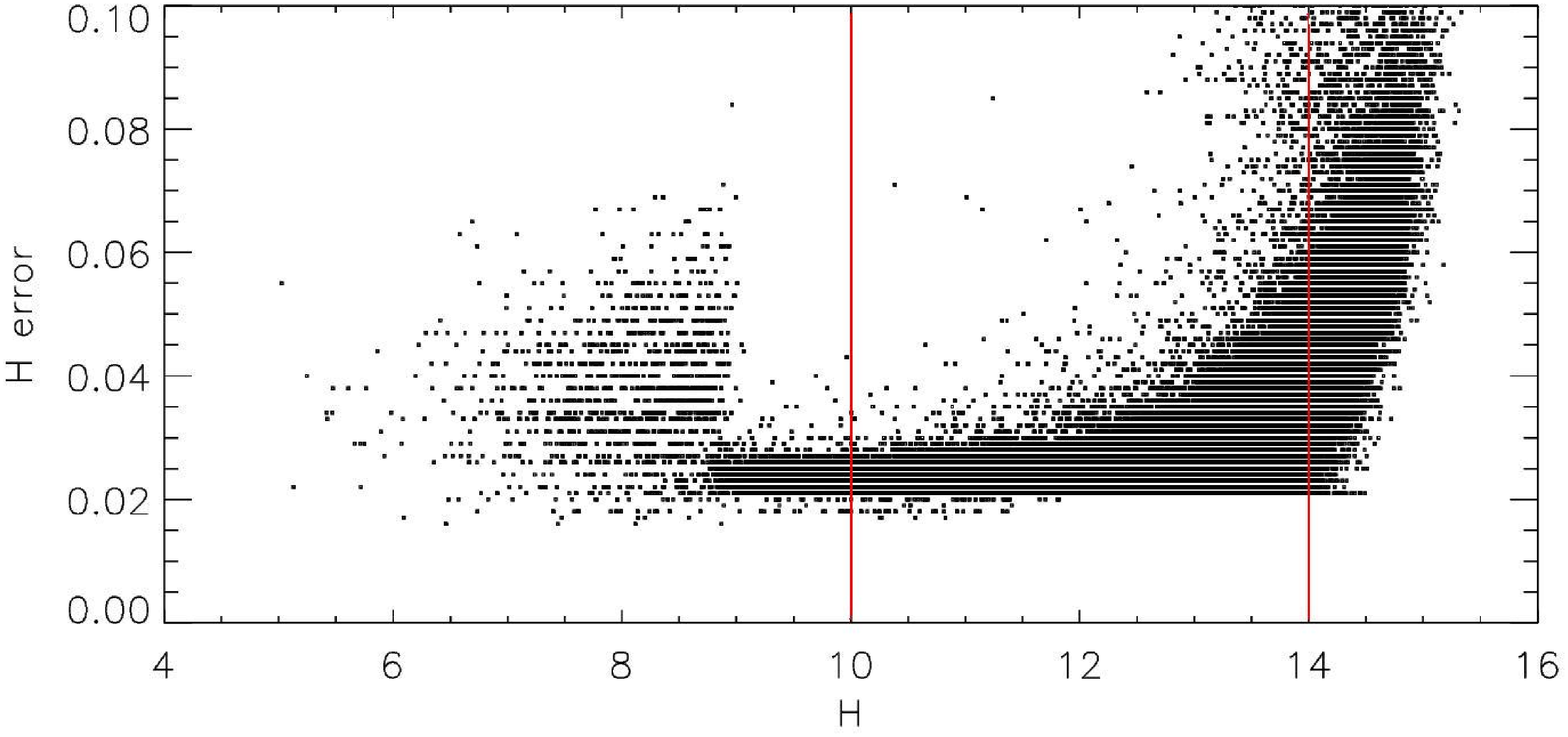}}
\resizebox{\hsize}{!}{\includegraphics{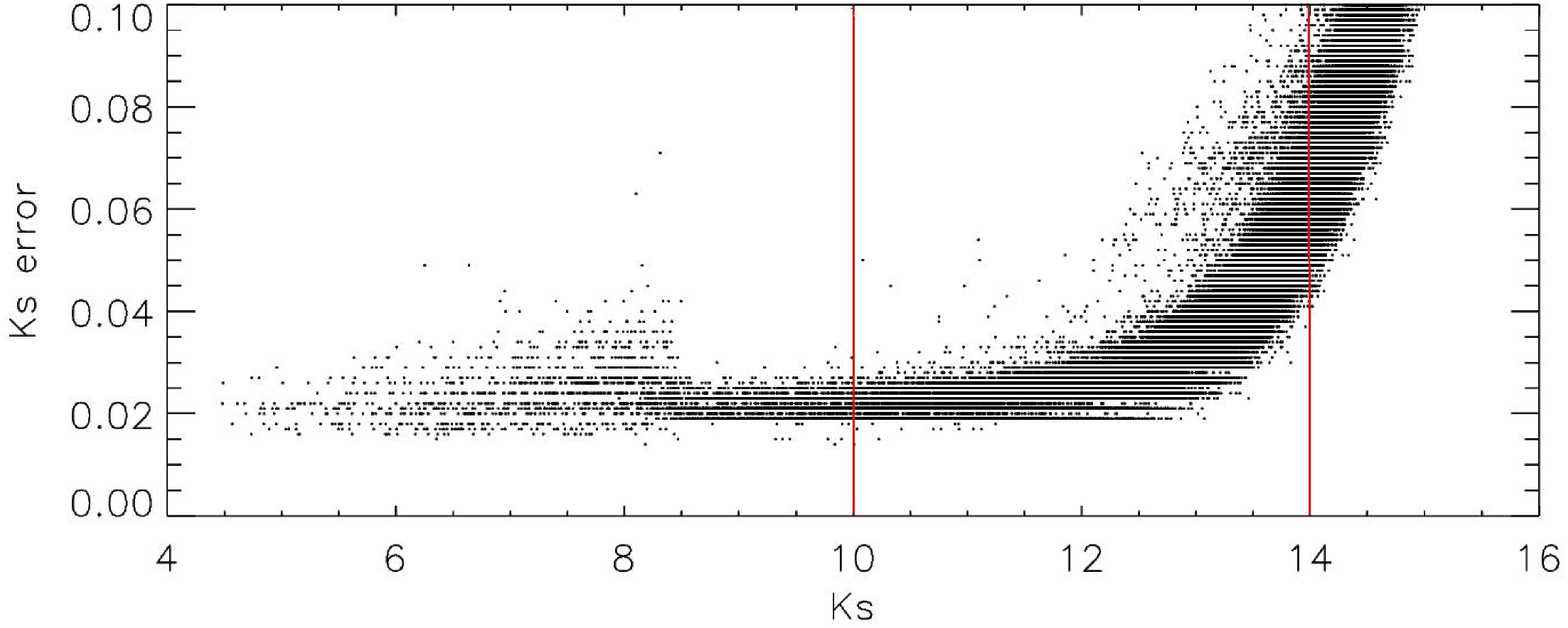}}
  \caption{A comparisson of magnitude versus error for the 2MASS $J$, $H$, $K_\mathrm{s}$ bands. Magnitude cuts are shown in red at $J$, $H$, $K_\mathrm{s}$ = $10.0$, $14.0$ as described in Sect. \ref{data}.}
\label{errmag}
\end{figure}

\section{Figures}
\begin{figure}[H]
\resizebox{\hsize}{!}{\includegraphics{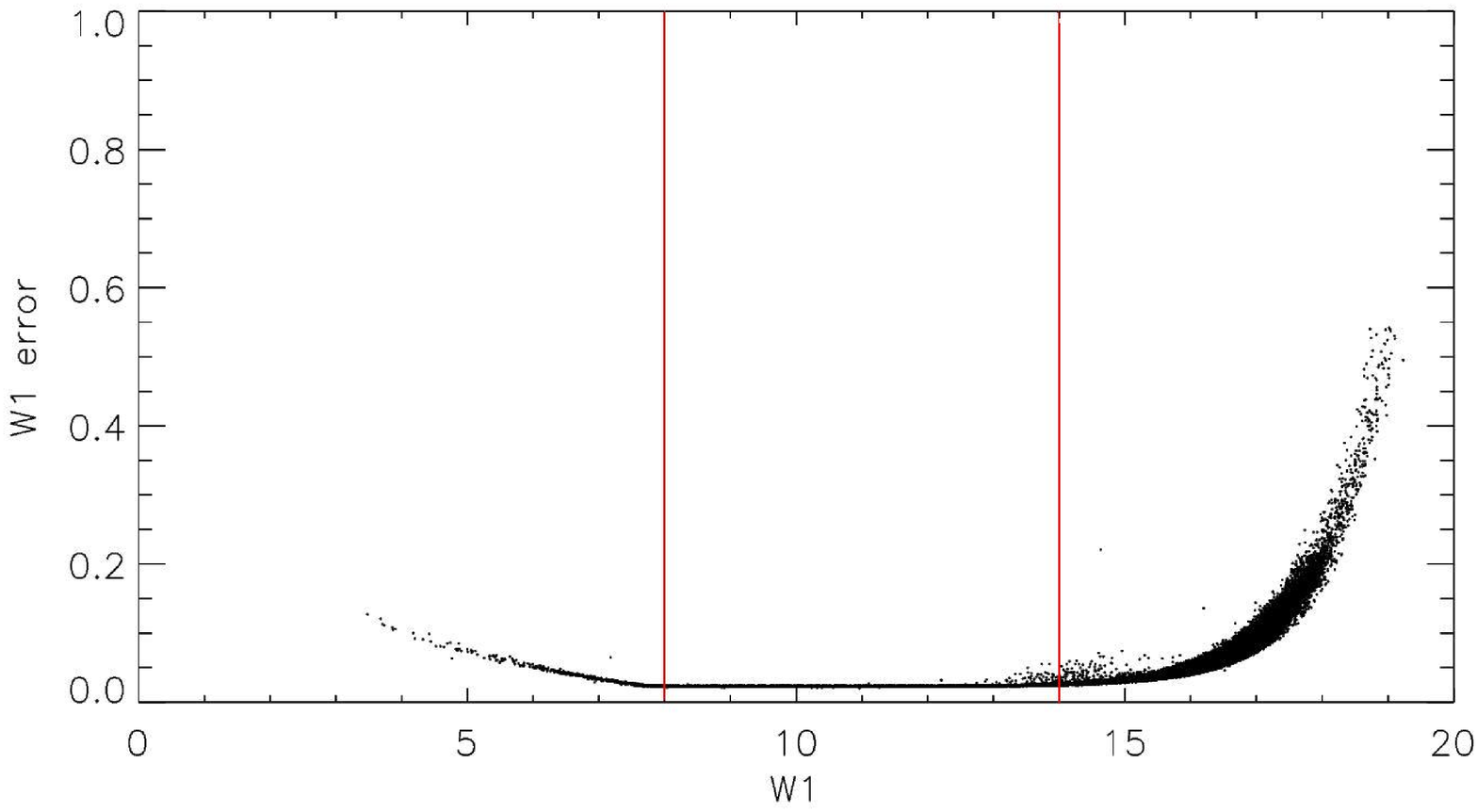}}
\resizebox{\hsize}{!}{\includegraphics{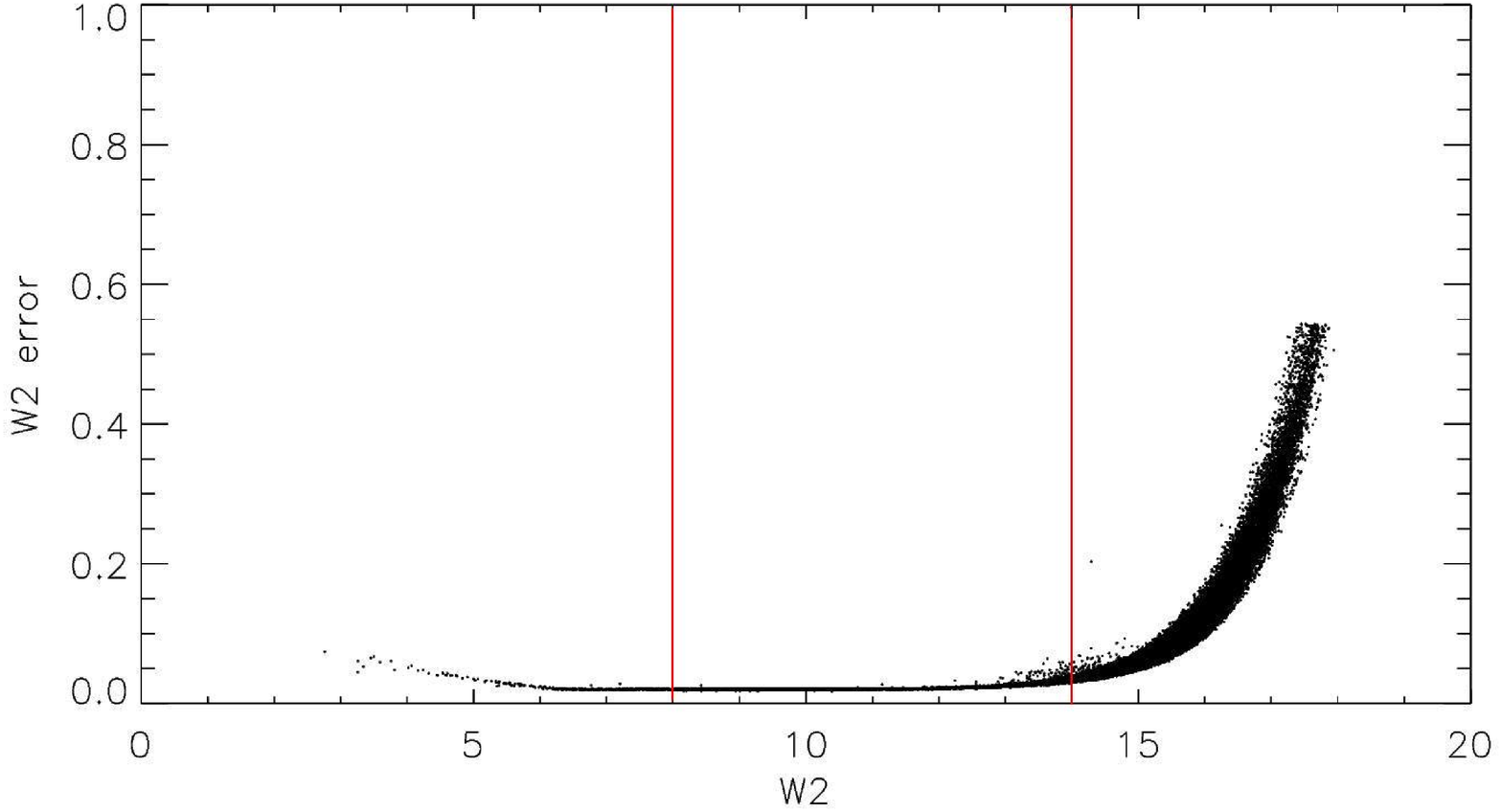}}
  \caption{A comparisson of magnitude versus error for the WISE $W1$ and $W2$ bands. Magnitude cuts are shown in red at $W1$, $W2$ = $8.0$, $14.0$ as described in Sect. \ref{data}.}
\label{errmag2}
\end{figure}

\end{document}